\documentclass[conference]{IEEEtran}
\IEEEoverridecommandlockouts

\usepackage{cite}
\usepackage{amsmath,amssymb,amsfonts}
\usepackage{algorithmic}
\usepackage{graphicx}
\usepackage{textcomp}
\usepackage{xcolor}
\usepackage{authblk}
\usepackage{multirow}

\def\BibTeX{{\rm B\kern-.05em{\sc i\kern-.025em b}\kern-.08em
    T\kern-.1667em\lower.7ex\hbox{E}\kern-.125emX}}

\begin{document}

\title{M3-CVC: Controllable Video Compression with Multimodal Generative Models\\}

\author{Rui Wan, Qi Zheng, Yibo Fan*\thanks{\hspace{-0.2cm}\rule{6cm}{0.6pt}\\* Corresponding author.}\\
\textit{State Key Laboratory of Integrated Chips and Systems} \\
\textit{Fudan University}\\
Shanghai, China \\
}
\maketitle

\begin{abstract}

Traditional and neural video codecs commonly encounter limitations in controllability and generality under ultra-low-bitrate coding scenarios. To overcome these challenges, we propose M3-CVC, a controllable video compression framework incorporating multimodal generative models. The framework utilizes a semantic-motion composite strategy for keyframe selection to retain critical information. For each keyframe and its corresponding video clip, a dialogue-based large multimodal model (LMM) approach extracts hierarchical spatiotemporal details, enabling both inter-frame and intra-frame representations for improved video fidelity while enhancing encoding interpretability. M3-CVC further employs a conditional diffusion-based, text-guided keyframe compression method, achieving high fidelity in frame reconstruction. During decoding, textual descriptions derived from LMMs guide the diffusion process to restore the original video's content accurately. Experimental results demonstrate that M3-CVC significantly outperforms the state-of-the-art VVC standard in ultra-low bitrate scenarios, particularly in preserving semantic and perceptual fidelity.
\end{abstract}


\section{Introduction}
In recent years, motivated by the need for efficient compression in bandwidth-constrained environments and emerging video applications, traditional video coding standards like H.264(AVC)\cite{H.264}, H.265(HEVC)\cite{H.265} and H.266(VVC)\cite{H.266} have been developed. However, these standards are built on rigid assumptions about content structure, resulting in constrained adaptability when confronted with increasingly complex video content\cite{zheng2024videoqualityassessmentcomprehensive}. In this context, deep learning-based video codecs have been proposed to dynamically learn data-driven representations, allowing for enhanced signal restoration capabilities. 

Deep learning-based video codecs can be broadly categorized into two types: general-purpose learned codecs and domain-specific codecs. General-purpose learned codecs are designed to compress various types of videos. Common Techniques employed include spatiotemporal contextual modeling\cite{DCVC-DC,DCVC-FM} and optimization utilizing generative models such as Generative Adversarial Networks (GAN)\cite{NVC-GAN}. General-purpose learned codecs exhibit considerable versatility, while they frequently exhibit elevated compression bitrates and operate as black-box systems, rendering their internal mechanisms opaque and challenging to interpret or control directly. Domain-specific codecs focus on compressing particular types of videos\cite{zheng2024unicornunifiedneuralimage}, such as human face\cite{gfvc} and body videos\cite{body_gvc}. Typical frameworks characterize video frames into symbolic representations including learned 2D keypoints\cite{dac} and 2D landmarks\cite{gfvc_2dlm}. Domain-specific codecs typically achieve higher efficiency in low-bitrate compression while offering interpretable features. Nonetheless, their applicability is limited. Currently, there are few neural video codecs\cite{CMVC} possessing both general applicability and controllability while achieving extremely low bitrates. To address this gap, it is imperative to enhance the extraction of spatiotemporal and semantic information from videos and leverage this information for efficient video reconstruction\cite{gvc}. The emergence of large multimodal models\cite{mllm} and conditional diffusion models\cite{cdm} presents significant opportunities for advancing these tasks, as they offer improved capabilities for integrating and utilizing complex information across modalities with strong priors.

Large multimodal models (LMMs)\cite{videollava,qwen-vl} offer significant advancements in the extraction of visual information from videos and images. Compared to traditional image-to-text or video-to-text models, LMMs are capable of performing in-depth analyses of objects and their intricate relationships within visual data.  Additionally, LMMs demonstrate superior interactive flexibility, enabling targeted information extraction and description generation based on user-defined prompts. These attributes underscore the efficacy of LMMs in providing efficient and controllable extraction of perceptual and semantic information from visual media. Conditional diffusion models (CDMs) exhibit substantial advantages over traditional generative models including GAN in conditional image and video generation. Embedded with extensive priors, conditional diffusion models facilitate a range of complex video generation tasks based on guiding conditions (textual descriptions, reference frames, etc.). These tasks encompass video/image generation\cite{videocrafter2,sd}, video prediction\cite{seine} and frame interpolation\cite{diffmorpher}.

\begin{figure*}[htbp]
    \centering
    \includegraphics[width=\textwidth]{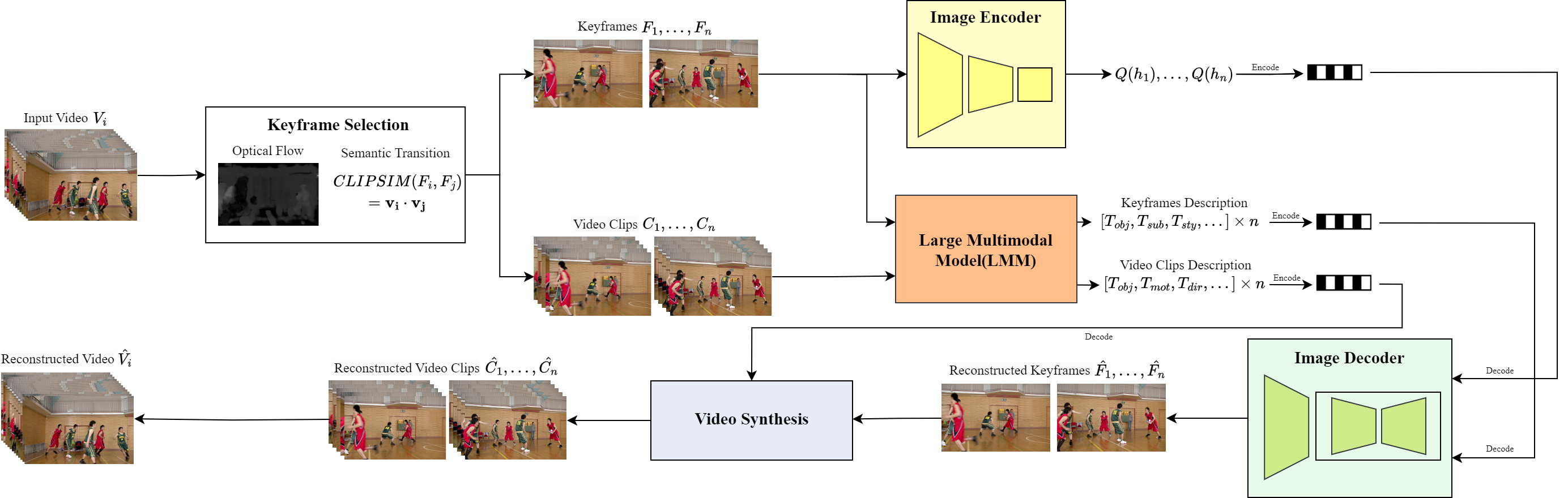}
    \caption{Overview of proposed M3-CVC framework}
    \label{system}
\end{figure*}

Drawing from the analysis above, we propose \textbf{M3-CVC}, a versatile, controllable, and ultra-low-bitrate video compression system by utilizing multiple LMMs and CDMs. As shown in Fig. \ref{system}, the system initiates with segmenting the video into several clips through keyframe selection. Following this, a hierarchical strategy is adopted for encoding and decoding keyframes and video clips. For each keyframe, semantic and perceptual information is extracted using LMMs and losslessly encoded, while dimensionality reduction and quantization are performed via a neural encoder to produce entropy coding as another part of the encoded bitstream. Upon decoding, we utilize text-to-image diffusion model to reconstruct the keyframe. For each video clip, the LMM extracts segment-level information, which is losslessly encoded consequently. On the decoder side, text description and its corresponding restored keyframe act as conditional input of video generation diffusion model, which reconstructs the video clip. The complete video is then assembled by concatenating all the reconstructed clips. This architecture facilitates ultra-low-bitrate video compression with high semantic and perceptual fidelity. Furthermore, by interpreting the textual output from the LMM, the compression process can be monitored, and fine-grained adjustments to keyframe and video encoding can be made dynamically by modifying the LMM prompts.

\section{Methodology}

\subsection{Keyframe Selection}\label{AA}
To retain the key visual information of a video, we employ a semantic-motion composite decision strategy for keyframe selection. We define two functions: The first function \( f_{\text{Clip}}(F_i, F_j) \), computes semantic features \( \mathbf{v}_i \) and \( \mathbf{v}_j \) using CLIP\cite{clip} model. It returns the cosine similarity between features. The second function, \( f_{\text{Raft}}(F_i, F_j) \), utilizes the optical flow prediction model RAFT \cite{raft} to calculate the magnitude of motion changes between the two frames.

Based on these two functions, we define a weighted sum function \( D(\cdot, \cdot) \) to determine whether the current frame should be selected as a keyframe. The function is expressed as follows:
\begin{align}
D(F_n,F_l) &=\lambda_1(1-f_{Clip}(F_n,F_l)) + \lambda_2f_{Raft}(F_n,F_l)-D_{th} \nonumber 
\end{align}
where \( F_l \) and \( F_n \) represent the previous keyframe and current frame respectively; \( \lambda_1 \) and \( \lambda_2 \) are adjustable parameters; \( D_{\text{th}} \) is a tunable decision threshold. If \( D(F_l, F_n) > 0 \), the frame \( F_n \) is selected as the new keyframe. The introduction of the function \( D(\cdot, \cdot) \) enables keyframe selection based on multiple video attributes. 

\subsection{Utilization of LMM}\label{BB}
\begin{figure}[htbp]
    \centering
    \includegraphics[width=0.45\textwidth]{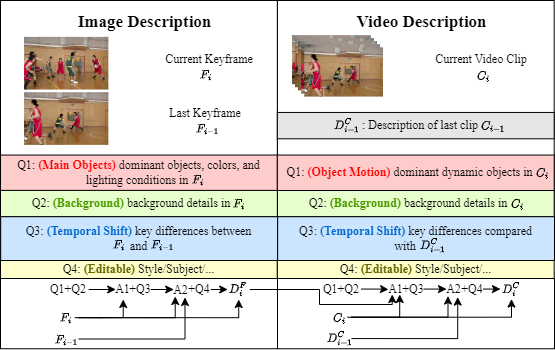}
    \caption{\centering{Multi-round dialogue-based visual information extraction strategy}}
    \label{prompt}
\end{figure}

Assuming that \( n \) keyframes \( F_1, \dots, F_n \) are selected, the video can be divided into \( n \) clips \( C_1, \dots, C_n \). For each pair of keyframe \( F_i \) and clip \( C_i \), we use pretrained Qwen-VL-Instruct-7B \cite{qwen-vl} to extract visual information. Inspired by interactive question answering strategy \cite{iqa} applied in multimodal learning, we designed a multi-round dialogue-based information extraction strategy to fully capture spatiotemporal information, as illustrated in Fig. \ref{prompt}. For keyframes, we first extract object and background information, then compare it with the preceding keyframe to obtain differential information, and finally extract trivial details and summarize them. The output \( D_i^F \) serves as the textual description of the keyframe. A similar dialogue strategy is employed for the video clips, where we further analyze the primary motion information based on the existing keyframe description \( D_i^F \), ultimately producing \( D_i^C \) as the textual description of the video clip. To improve transmission efficiency, we applied LZW encoding for lossless compression on \( D_i^F \) and \( D_i^C \).

\subsection{Keyframe Compression}\label{CC}
\begin{figure}[htbp]
    \centering
    \includegraphics[width=0.47\textwidth]{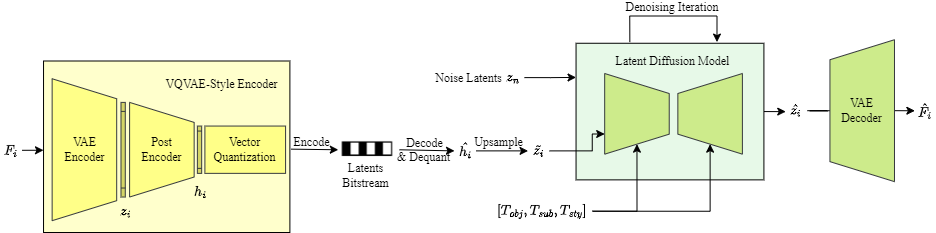}
    \caption{\centering{Overview of generative image codec for keyframe compression}}
    \label{enc-dec}
\end{figure}

To further reduce the video compression bitrate and ensure the perceptual and semantic fidelity of reconstructed keyframes, inspired by \cite{lic-cdm,perco}, we employ an image codec based on VAE and conditional latent diffusion models for encoding and decoding each keyframe \( F_i \in \mathbb{R}^{3 \times h \times w} \) (\( i=1,\dots,n \)). The framework of the compressor is shown in Figure \ref{enc-dec}. 
The encoder side follows a VQVAE-style \cite{VQVAE} neural architecture. First, the pretrained VAE encoder projects \( F_i \) from the pixel space into the latent space variable \( \mathbf{z_i} \in \mathbb{R}^{ch_l \times h_l \times w_l} \). Inspired by the ELIC encoder \cite{elic}, a post-encoder module is introduced to encode the latent representation \( \mathbf{z_i} \) to generate the distribution parameters of the latent variables, producing the encoded representation \( \mathbf{h_i} \in \mathbb{R}^{ch_p \times h_p \times w_p} \). Finally, the representation \( \mathbf{h_i} \) is quantized based on the vocabulary \( \mathbf{V} \in \mathbb{R}^{ch_p \times l_v} \):
\begin{equation*}
\begin{split}
k^*(u, v) &= \underset{k \in \{1, \dots, l_v\}}{\arg\min} \, d(\mathbf{h_i}(u, v), \mathbf{v_k}) \\
\mathbf{K^*} &= \left( k^*(i, j) \right)_{i=1, \dots, h_p}^{j=1, \dots, w_p}
\end{split}
\end{equation*}
where $d(\cdot,\cdot)$ represents the Euclidean distance calculation.
$\mathbf{K^*}$ records the discrete quantization results of $\mathbf{h_i}$, and is entropy coded as a component of the compressed bitstream.

The decoder side utilizes a conditional diffusion model to reconstruct keyframes. After entropy decoding of \(\mathbf{K^*}\), it is reconstructed into a tensor of size \( ch_p \times h_p \times w_p \) by querying the vocabulary:
\[
\mathbf{\hat{h_i}}(u, v) = \mathbf{v_{k^*(u,v)}} \quad \forall (u, v) \in \{1, \dots, h_p\} \times \{1, \dots, w_p\}
\]
Subsequently, \(\hat{h_i}\) is upsampled to obtain the reconstructed latent variable \(\hat{z_i} \in \mathbb{R}^{\text{ch}_l \times h_l \times w_l}\). The conditional denoising process utilizes \(\hat{z_i}\) as  conditional variable and the keyframe description \(D_i^F\) as textual guidance. We select Stable Diffusion \cite{sd} as conditional denoising model. The iterative denoising process can be expressed as follows:
\[
z_{d,T} \sim \mathcal{N}(\mathbf{0}, \mathbf{I})
\]
\[
z_{d,t-1} = \frac{1}{\sqrt{\alpha_t}}\left( z_{d,t} - \frac{1 - \alpha_t}{\sqrt{1 - \bar{\alpha}_t}} \epsilon_{\theta}(z_{d,t}, t, \hat{z_i}, D_i^F) \right) + \sigma_t \mathbf{z}
\]
where ${\alpha}_t,\bar{\alpha}_t,\sigma_t$ are tunable factors, \(z_{d,t}\) is the denoising variable at timestep \(t\), \(\epsilon_{\theta}\) represents the denoising model, \(T\) is the total number of denoising steps, and \(\mathbf{z}\) is standard Gaussian noise. After denoising, the reconstructed latent variable \(z_{d,0}\) is processed using a pre-trained VAE decoder to obtain the reconstructed keyframe \(\hat{F_i}\).

\subsection{Video Reconstruction}\label{CC}
After obtaining the reconstructed video frames \(\hat{F_1}, \dots, \hat{F_n}\), along with the clip description \(D_1^C, \dots, D_n^C\) acquired during decoding, the \(n\) video clips can be decoded and reconstructed. We select the pre-trained video generation diffusion model SEINE \cite{seine}  for video reconstruction. We design two video reconstruction mode named "prediction" and "interpolation" respectively. In the prediction mode, the model takes the current keyframe \(\hat{F_i}\) and video clip description \(D_i^C\) as input, and if the total length of the video clip is \(m\) frames, it predicts the subsequent \(m-1\) frames. In the interpolation mode," the model incorporates the next keyframe \(\hat{F_{i+1}}\) as an additional input along with the current keyframe and video clip description, and similarly interpolates \(m-1\) frames. Both modes allow adjustments to frame rate and output resolution (up to $512 \times 512$) to accommodate different video reconstruction tasks. After completing the reconstruction task, \(n\) reconstructed video clips \(\hat{C_1}, \dots, \hat{C_n}\) are obtained, and concatenating these \(n\) clips produces the final decoded video \(\hat{V} = \{\hat{C_1}, \dots, \hat{C_n}\}\).

\section{Experiments}
\subsection{Experimental Setup}\label{XX}
\textbf{Training Details} Instead of fine-tuning pretrained LMM Qwen-VL-Instruct-7B or video generation model SEINE, our training work mainly focuses on the keyframe codec to fully utilize text descriptions for generating high-fidelity keyframes. We first perform pretraining of the keyframe encoder using the Commit loss commonly used in the VQVAE-style encoder:
\[
L_{\text{commit}} = \| z - \text{sg}(z_q) \|_2^2
\]
where $z$ is the continuous latent variable, $z_q$ is the closest quantized vector, and $sg(\cdot)$ is the stop-gradient operator. We set the vocabulary size of the keyframe encoder to a fixed value of 256, and the output dimensions of the Post Encoder \(h_p \times w_p\) are set to $32 \times 32$ and $64 \times 64$ as two options. Encoders with different options are trained separately. Subsequently, we fine-tune linear layers of the denoising network in image decoder based on the two pretrained encoders respectively, with the fine-tuning loss function as follows:
\[
L_t = \mathbb{E}_{\mathbf{z}_t, \epsilon, c} \left[ \|\epsilon - \epsilon_\theta(\mathbf{z}_t, t, c)\|^2 \right]
\]

\textbf{Datasets} We utilize the MSR-VTT dataset \cite{msr-vtt} as the training dataset. All training videos are cropped and resized to $512 \times 320$, with frame rate adjusted to 8 fps. Subsequently, several keyframes are extracted from each training video using method based on semantic difference (setting $\lambda_2=0$ in \ref{AA}). Pretrained LMM Qwen-VL-Instruct-7B \cite{qwen-vl} is utilized to extract visual information, ultimately producing a set of text-image pairs for end-to-end training of the image codec. As for test dataset, we select UVG \cite{uvg}, MCL-JCV \cite{mcl-jcv}, HEVC Class B and C to evaluate the performance of traditional codecs, learned codecs, and our proposed framework. All test videos are preprocessed with the same procedure as applied to the training videos.

\begin{figure*}[htbp]
    \centering
    \includegraphics[width=\textwidth]{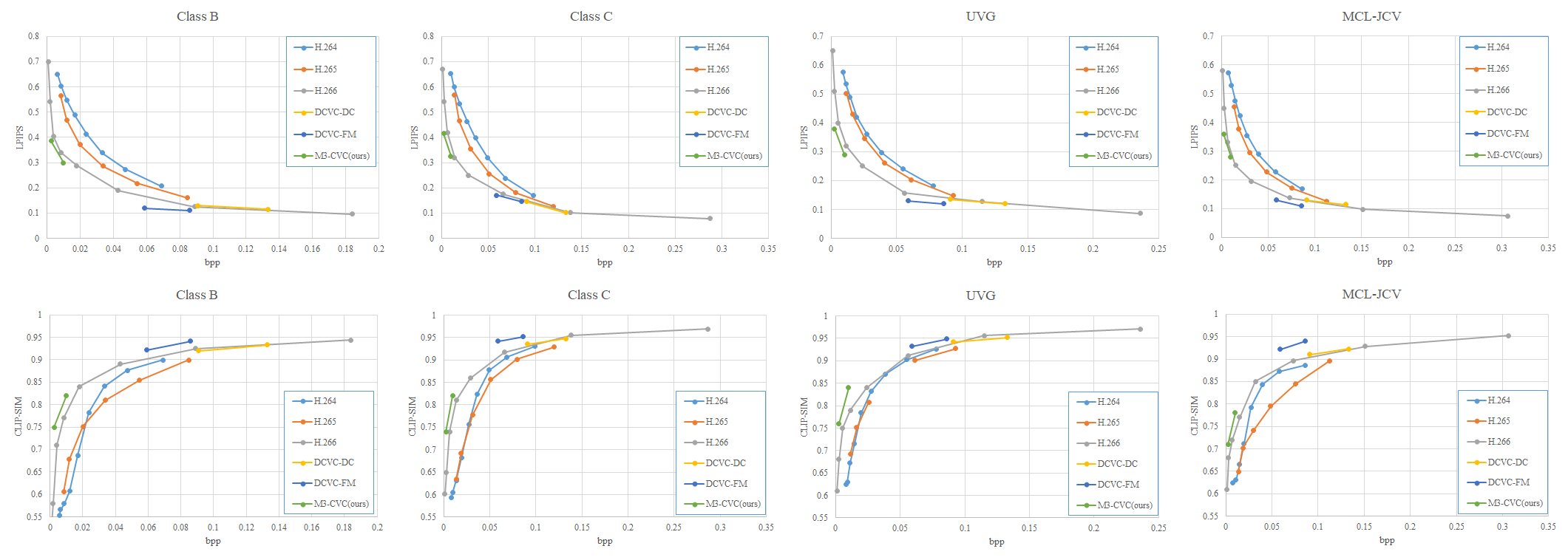}
    \caption{The R-D performance evaluation with LPIPS and CLIP-sim performed on HEVC Class B, Class C, UVG, and MCL-JCV Datasets.}
    \label{rd}
\end{figure*}

\textbf{Metrics} We assess video compression efficiency using bits per pixel (bpp) to evaluate the rate, while distortion is measured with CLIP-sim \cite{clip} and LPIPS. CLIP-sim evaluates the semantic similarity, while LPIPS measures the perceptual similarity between the reconstructed and original videos.

\subsection{Quantitative and Qualitative Analysis}\label{YY}
\begin{figure}[htbp]
    \centering
    \includegraphics[width=0.48\textwidth]{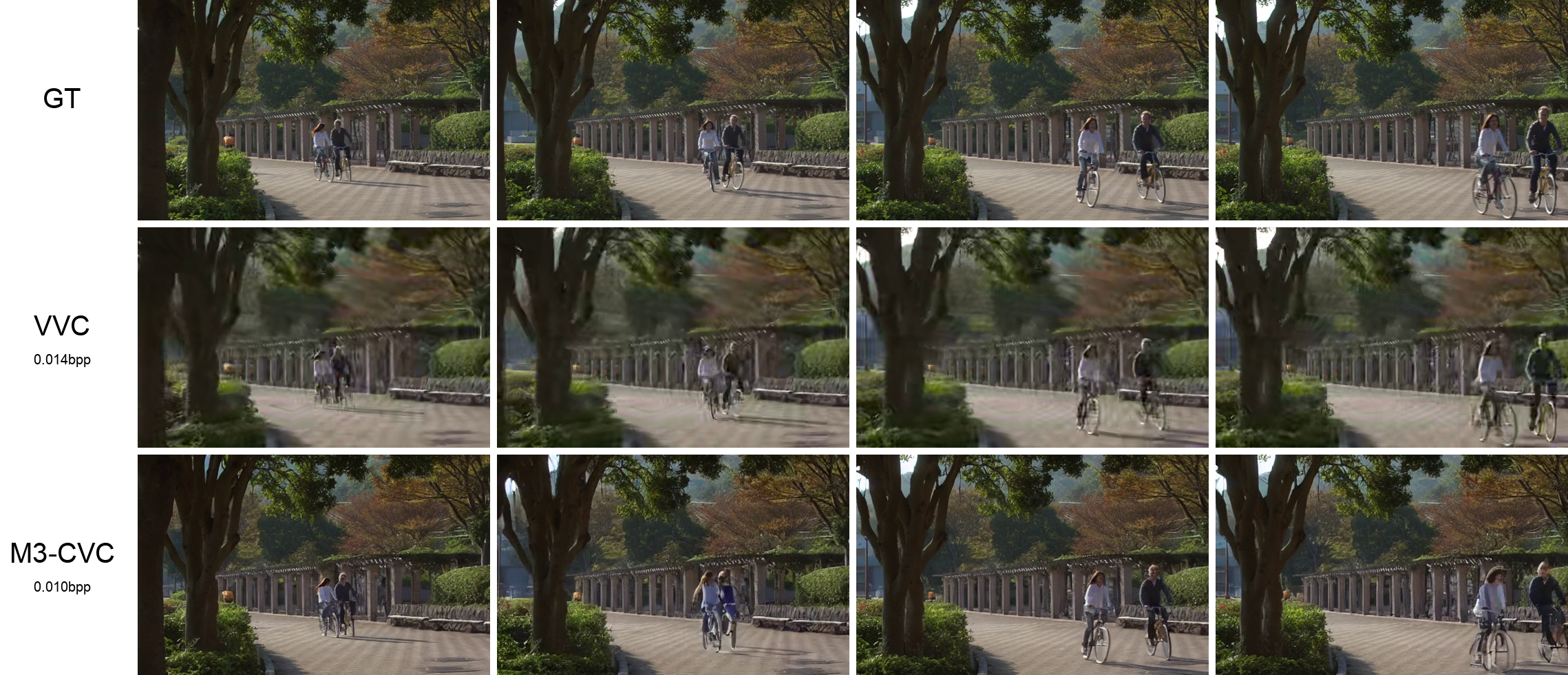}
    \caption{Visual quality comparison between ground truth Video, VVC and M3-CVC}
    \label{vis}
\end{figure}

To comprehensively evaluate the performance of the proposed M3-CVC framework, we conduct comparisons with both traditional codecs (H.264, H.265, and H.266) and learned codecs (DCVC-DC and DCVC-FM). For H.264 and H.265, FFmpeg x264 and x265 encoders are used with "veryslow" setting. For H.266, we utilize VTM-17.0 under low-delay P configuration with QP values of {63, 58, 53, 48, 43, 37, 32, 27}. Based on the above settings, we conduct R-D performance experiments, and the resulting R-D curves and visual results are shown in \ref{rd} and \ref{vis} respectively. It is evident that M3-CVC outperforms the state-of-the-art VVC standard, achieving substantial bitrate savings. Notably, M3-CVC excels under extremely low bitrates, where its performance significantly surpasses that of traditional codecs. Furthermore, in terms of semantic fidelity, M3-CVC demonstrates particularly strong results, ensuring higher preservation of video content meaning while maintaining competitive performance compared to both DCVC-DC and DCVC-FM.

\subsection{Ablation Study}\label{ZZ}

To validate the effectiveness of the methods employed in our framework, we conduct a series of ablation experiments using VTM-17.0 as the anchor. These experiments are performed on three video sequences from the HEVC Class B dataset (\textit{Kimono1}, \textit{ParkScene}, and \textit{Cactus}). We evaluate the R-D performance using LPIPS and compute the BD-rate savings for each sequence. The overall BD-rate saving results are obtained by averaging the individual savings across all test sequences.

For keyframe selection strategy, we perform three comparative tests: fixed-interval keyframe extraction, semantic threshold-only and joint semantic-motion threshold. The results demonstrate that with an equal number of keyframes, the method where both kinds of thresholds are non-zero achieves superior R-D performance. Regarding image information extraction, we compare two approaches: having the LMM describe all information in a single step versus employing a multi-turn dialogue strategy. The results confirm the advantage of our dialogue strategy. For video generation modes, we compare interpolation mode and prediction mode, showing that interpolation mode, which incorporates additional visual signal information, yields significantly better performance.

\vspace{-5pt}

\begin{table}[htbp]
    \centering
    \caption{Ablation Studies of M3-CVC Framework}
    \label{abl}
    \begin{tabular}{|c|c|c|}
        \hline
        Models & Settings & BD-Rate(\%) \\
        \hline
        \multirow{3}{*}{\centering Keyframe Selection Strategy} 
                           & Fixed-interval & -13.4 \\\cline{2-3}
                           & $\lambda_1\neq 0,\lambda_2=0$  & -9.7 \\\cline{2-3}
                           & $\lambda_1\neq 0,\lambda_2\neq0$               & -18.9 \\\cline{2-3}

        \hline
        \multirow{3}{*}{\centering LMM Dialogue Turns} 
                           & 1                 & -9.4 \\\cline{2-3}
                           & 2                 & -15.2 \\\cline{2-3}
                           & 4                 & -19.5 \\\cline{2-3}
        \hline
        \multirow{2}{*}{\centering Video Reconstruction Mode} 
                           & Interpolation               & -20.4 \\\cline{2-3}
                           & Prediction            & 22.1 \\\cline{2-3}

        \hline
    \end{tabular}
\end{table}

\subsection{Latency Performance Evaluation}

We test the processing latency of M3-CVC on a single NVIDIA RTX 3090 GPU. Meanwhile, we also measure the processing latency of VTM-17.0 under low-delay P configuration on Intel Xeon Gold 6230 CPU with 40 cores as a comparison. The results are presented in Table \ref{tab:latency-comparison}. The experiments are conducted on \textit{ParkScene} video sequence from HEVC Class B dataset.

\vspace{-5pt}
\begin{table}[h!] 
\centering
\renewcommand{\arraystretch}{1.2} 
\setlength{\tabcolsep}{2pt}       
\caption{Latency Comparison of M3-CVC and VTM-17.0} 
\begin{tabular}{|c|c|c|c|}
\hline
\textbf{Item}      & \textbf{Latency / s} & \textbf{Item}  & \textbf{Latency / s} \\ \hline
{M3-CVC}           &        219.2               & VTM-17.0       &      1542.8                 \\ \cline{1-4}
Encoder           &        76.7               & Encoder       &      1519.7                 \\ \cline{1-4}
Decoder           &        142.5               & Decoder       &      23.1                 \\ \hline
\end{tabular}

\label{tab:latency-comparison}
\end{table}

The results demonstrate that M3-CVC outperforms VTM-17.0 in overall latency, particularly during encoding, due to GPU-accelerated generative models. While decoding latency is higher in M3-CVC, a common issue with diffusion-based codecs \cite{CMVC}, the method shows promise for competitive video compression, especially when encoding latency is prioritized.

\section{Conclusion}
In this paper, we introduce M3-CVC, a framework designed for general-purpose ultra-low-bitrate video compression. M3-CVC realizes controllable video information extraction by leveraging large multimodal models with specifically-designed dialogue strategies, while achieves high-fidelity video reconstruction with pretrained conditional diffusion models designed for image and video generation. Experimental results demonstrate that M3-CVC shows significant performance improvements compared to state-of-the-art technologies such as VVC, particularly under extremely low bitrate conditions.

\bibliographystyle{unsrt} 

\vspace{12pt}

\end{document}